%% file: main.tex
\documentclass[conference]{IEEEtran}

\usepackage{textcomp}
\usepackage{comment}
\usepackage{ifthen}
\usepackage{subcaption}
\usepackage{booktabs}
\usepackage{footnote}
\usepackage{graphicx}
\usepackage{paralist}
\usepackage{enumitem}
\usepackage{float}
\usepackage{color}
\usepackage{xspace}	
\usepackage{flushend}
\usepackage{balance}
\usepackage{enumitem}
\usepackage{marvosym}
\usepackage{hyperref}

\def\BibTeX{{\rm B\kern-.05em{\sc i\kern-.025em b}\kern-.08em
    T\kern-.1667em\lower.7ex\hbox{E}\kern-.125emX}}

\newboolean{authnotes}

\ifthenelse{\boolean{authnotes}}
{
	\newcommand{\amy}[1]{\footnote{{\bf Amy: #1}}}
	\newcommand{\rajeev}[1]{\footnote{{\bf Rajeev: #1}}}
}
{
	\newcommand{\amy}[1]{}
	\newcommand{\rajeev}[1]{}	
	\newcommand{\michele}[1]{}
}

\newcommand{\fakeparagraphnodot}[1]{\vspace{0mm}\noindent\textbf{#1}}
\newcommand{\fakeparagraph}[1]{\fakeparagraphnodot{#1}}

\newcommand{\wurbench}{\texttt{WURBench}\xspace}
\newcommand{\lpwur}{LP-WUR\xspace}
\newcommand{\wur}{WUR\xspace}
\newcommand{\wurs}{WURs\xspace}

\begin{document}

\title{WURBench: Toward Benchmarking Wake-up Radio-based Systems}

\author{\IEEEauthorblockN{Rajeev~Piyare, Amy~L.~Murphy }
	\IEEEauthorblockA{Fondazione Bruno Kessler\\
		Via Sommarive 18, 38123, Povo, Trento, Italy\\
		Email: \{piyare, murphy\}@fbk.eu}
}

\maketitle

\input{abstract.tex}


\input{intro.tex}

\input{hardware.tex}

\input{system.tex}

\input{conclusion.tex}

\balance
\bibliographystyle{IEEEtran}
\bibliography{bib}

\end{document}

%% file: abstract.tex
\begin{abstract}
	

The performance of wake-up radios must be clearly measured and understood while designing and developing robust, dependable, and affordable systems, considering both
benefits and shortcomings.  State-of-the-art WURs display
significant diversity in their architecture, processing capability, energy
consumption, and receiver sensitivity. Standard methodologies for benchmarking
are crucial for quantitatively evaluating the performance of this emerging
technology, however, currently, no accepted standard for such quantitative
measurement exists. Further, there is no consensus on what objective
evaluation procedures and metrics should  be used to understand the
performance of  whole systems exploiting this technology.  This lack of
standardization has prevented  researchers from comparing results and
leveraging previous work that could otherwise avoid duplication and speed up
the validation process. This paper leads toward an evaluation framework, a \emph{benchmark},
to enable accurate and repeatable profiling of WUR-based systems, leading to more
consistent and therefore comparable evaluations for current and future systems.

%
 
\end{abstract}


%% file: intro.tex
\section{Introduction}
\label{sec:introduction}
With the proliferation of the Internet of Things devices and the seemingly
endless connections among people, the demand for reliable and long-lasting
devices is becoming critical. IoT devices are expected to be network
connected at all times, even while simply waiting for an event to happen. Nevertheless, conventional solutions have significant costs even in this standby state, consuming on the order of a few milliwatts and reducing the useful device lifetime. 

One emerging solution to extend battery life is the incorporation of a
Low-Power Wake-up Radio (\lpwur)~\cite{rajeevsurvey}. A \lpwur consumes only micro-watts of power, usually below 100~$\mu$W,
and is continuously on, listening for a trigger either on the same
channel used for data communication or on a dedicated, out-of-band
channel. Once a trigger is received, the \lpwur activates the primary device,
which can otherwise remain in a deep sleep state, saving energy.  The extreme
low power constraints of WUR limit the receiver complexity, modulation scheme,
and thus the overall receiver sensitivity of WUR designs, reducing the
effective communication
range. 

The potential for \lpwur is rapidly expanding into domains such
as Wi-Fi access points, low-power wide area sensor networks~\cite{aoudia2018long}, and wildlife
monitoring. As \lpwur moves towards a widely used practical technology,
simulations and experimental deployments must be performed  to validate proposed hardware designs, network architectures, and wireless communication protocols.


\fakeparagraph{Problem.}  Current state-of-the-art \wur prototypes, most of
which are custom in-lab designs, display significant diversity in their
architectures, processing capabilities, energy consumptions, and receiver
sensitivities~\cite{rajeevsurvey}.  Given such diversity in platform
characteristics, choosing the right prototype and protocol for a specific
application scenario is challenging.  Nevertheless, the available
communication protocols are evaluated in restricted settings with ad-hoc
experiments and without comparison to competing approaches, making it almost
impossible to identify the appropriate protocol for a given prototype radio in
a specific application.  Moreover, results of experiments performed in
different settings (e.g., topology, traffic pattern, interference) using
different metrics may not hold for others. Currently, there exist no fixed set
of accepted testing methods, parameters or metrics to be applied to a WUR
based system under evaluation. This lack of standardization significantly
increases the difficulty to develop new systems and/or apply existing
technologies in novel domains.

\fakeparagraph{Solution.}  To overcome these disparities, we identify the key
parameters of a new evaluation methodology, \emph{\wurbench}, offering a step
toward enabling accurate and repeatable profiling of WUR-based systems for IoT
applications. Further, we outline the issues that must be addressed before
full WUR benchmarking can become a reality.

The concept of benchmarking is not new and has been applied to areas such as
wireless networking~\cite{2018iotbench} and CPUs~\cite{cpu}
to compare performance results. Recently,
IoT-Connect~\cite{iot-connect}, an Industry-Standard Benchmark for Embedded
Systems has been introduced  to evaluate micro-controllers with various
connectivity interfaces such as Bluetooth, LoRa, and WiFi. 
A benchmark typically outlines a set of specifications to follow when
evaluating the performance of a system, making experiments
repeatable and results directly comparable. These specifications include the
definition of the parameters for the experimental
setup and output metrics reflecting  the performance of the benchmarked
system. Benchmarking WUR is non-trivial as this not only requires
evaluating the WUR prototypes and protocols, but also measuring or modeling
the wireless environment such as interference sources that have the potential
to significantly affect system performance.

\fakeparagraph{Goals.}
The main goal of \wurbench is to outline a benchmarking framework that will:

\begin{enumerate}[label=(\roman*)]
	\item provide a set of recommended practices for performance evaluation.
	
	\item offer reliable indicators in terms of key  performance \emph{metrics},
    \emph{parameters}, and \emph{tools} for researchers to test and fairly
    compare new solutions against existing ones or baselines when
    implementations are not publicly available. WUR hardware designers can
    also utilize this framework to benchmark devices against competitors.
    
    \item facilitate a \emph{repeatable}  test environment for WUR-based systems.
\end{enumerate}


%% file: hardware.tex
\section{Hardware Micro-benchmarking}
\label{sec:hardware}

The design phase of \wur-based low-power networking starts with choosing an
appropriate \wur prototype and performing a series of hardware specific
micro-benchmarks. Micro-benchmarks are small test applications that iterate
through the states of the component being tested e.g., radio, MCU, LEDs. As
\wurs are mostly custom designed, micro-benchmarking enables the
identification of possible performance bottlenecks at the architecture level,
allowing hardware designers to compare and assess design trade-offs.  Most
often these micro-benchmarks are conducted in an ad-hoc fashion 
limiting comparability. Here, we seek to provide a
well-defined structure, defining ``\emph{what to measure}'' and
``\emph{recommended practice}'' for measuring, allowing results to be compared
across prototypes and to verify the fidelity of the test platform.

\subsection{What to measure?} 
The first step is the definition of the \emph{metrics}; hardware performance
measured in terms of a set of quantitative variables of interest. At an
abstract level, the metrics defined here are hardware-agnostic making them
comparable across various \wur prototypes.

\begin{enumerate}[label=(\roman*)]
\item \emph{Communication range}: the achievable distance between the
  endpoints to establish a baseline for the performance of a given
  connection. For instance, evaluating transmit power vs. range is 
  important for \wur deployments.
	
	\item \emph{Successful wake-up rate}: is the communication reliability of
    the \wur module  measured in terms of the frame loss, which is the
    fraction of triggers sent by the sender over those successfully received at
    the receiver. This metric depends on the effective communication range 
    and the strength of the wake-up signal.
	
	\item\emph{Energy consumption}: right now there is no common way to
    universally benchmark the energy efficiency of the \wur and one must
    either independently benchmark or rely on the information provided in the literature. Energy consumption, for instance, may refer to the average power consumption of the \wur in the continuous channel monitoring state,  which is of greater importance for the IoT applications. Depending on the nature of the micro-benchmarking, this may also refer to the energy consumed by the \wur while executing different tasks for e.g., signal transmission and processing cost.  
\end{enumerate}

\subsection{Recommended practice}
To correctly measure the defined metrics, it is therefore mandatory to layout
the steps one needs to follow while conducting these experiments, specifying
the experimental parameters when characterizing these micro-benchmarks. The
\emph{parameters} are the configurations that allow controlling the execution
of the micro-benchmark. This is a critical piece of the evaluation as
comparing WURs without outlining all the configurations bring into
question the soundness of the comparison. The main parameters are identified
as:

\begin{itemize}
\item \emph{Physical layer (PHY) settings}: Most \wurs support various PHY
  settings that include bit rates, transmission power, and
  modulation. Often, some of these parameters are not identified, making it
  difficult to directly compare the prototypes, as seen
  in~\cite{rajeevsurvey}.  Therefore, the configurations used in each trial
  must be reported with the results. 

\item \emph{Antenna orientation}: The radiation patterns of the antenna
  determine the performance of wireless devices~\cite{antenna}.  It thus
  becomes important to state the antenna orientation and type in combination
  with frequency and transmit power.
	 	
\item \emph{Trial duration}: wireless links change over time due to subtle
  changes in environmental conditions. Therefore, evaluation should be spread
  throughout a 24~hour period and long duration experiments should be
  preferred. 

	
\item \emph{Firmware}: The firmware version being used in the tests as well as
  any functions that have been disabled, should be reported together with the
  results.
	
\item \emph{Environment}: The micro-benchmarking process is incomplete without
  describing the characteristics of the set-up environment that can be either
  based on real-life use case or artificial test environments. For a fair
  comparison, describing and documenting the scenarios for later analysis is
  critical.  The test setups can be divided coarsely into two categories:
	\begin{inparaenum}[\em i)]
		\item shielded: where cabling or RF shielding techniques are used to attenuate external signals and noise.
		\item open-air: environments that mimic the actual use case for the \wur such as indoor or outdoor environment with line-of-sight and non-line-of-sight.
	\end{inparaenum}
	
\end{itemize}



%% file: system.tex
\section{Benchmarking System as a Whole}
\label{sec:system}


While it is important to isolate functions for performance testing, benchmarking the software separately may not reveal all the functionalities or vulnerabilities of a system.  As software alone may not be able to capture underlying hardware behavior such as interrupts, timings, and design flaws, a complete-as-possible performance test must be conducted. In other words, benchmarking the whole system including the hardware and software interaction provides the most realistic evaluation.
For WSN systems, validation of the whole system is carried out using both,
simulations and testbeds, following these key steps: defining the application
scenario, choosing or implementing a communication protocol, conducting a
large set of experiments, measuring the performance in terms of defined
metrics, and comparing the results. This seemingly simple task of benchmarking
is surprisingly challenging for \wur networks as it requires an evaluation of
the entire system to capture real-world operational conditions. This requires
a large number of experiments that can be tedious and error-prone.  Moreover,
the complexity of this evaluation is compounded by the lack of control over
experimental conditions and lack of evaluation tools. Furthermore, results
obtained from ad-hoc experiments are difficult to compare with the results
gathered from different wireless testbeds and simulations, hindering
repeatability.  In this section, we focus on enabling benchmarking of
\wur-based systems as a whole by presenting a set of testing methods,
application scenarios, parameters, and metrics to be applied to a protocol
under test either using testbeds or simulation tools.

\subsection{Application settings}
The first critical element to define for a system as a whole is the expected
environment in which it will be exploited, as this defines many of they key
environmental parameters that influence a deployment. For example, the size of
the area to be covered and the density of measurement points affect the
network topology. Further, the application needs often direct the choice of
the protocol, for example data collection favors unidirectional focus while
control systems emphasize latency. These choices must be outlined clearly, as
they help focus the applicable metrics which we address in the following
section.

\subsection{Metrics} 
Next, we define key system performance and techno-economic metrics that 
we consider in the definition of \wurbench.
These can be divided  into four categories:
\begin{enumerate}[label=(\roman*)]
	\item \emph{Power consumption}: 
	computed using the amount of time a node keeps its radio on in different states such as Receive (RX), Transmit (TX), Idle, and Sleep. The consumption should also consider duty cycle patterns of both the radios to detect even small deviations that may have a substantial effect on the device lifetime in real deployments. 
	
	\item \emph{Reliability}: 
	defined as the fraction of application data packets successfully received over those sent. This is an indication of the level of service provided to applications in delivering sensed data, especially relevant when WUR is considered an option for safety-critical systems.

	\item \emph{Latency}: 
	defined as the end-to-end packet delivery delay from the time of generation to reception.
	
	\item \emph{Cost}: 
	one limitation of WUR comes from inherently low-power demand in continuous
  listening mode, which results in a limitation on the feasible distance
  between two devices. As such, the WUR needs to be combined with another
  system, typically with a high power node resulting in a dense network. The
  relative cost of replacing a standard node with a WUR-based one might incur
  additional cost. The system cost should, therefore, be calculated not only
  for the main sensor node but also for the extra \wur hardware.


\end{enumerate}

\subsection{Evaluation mechanisms}
Before proceeding with field tests, simulations and testbeds are the main tools for performance analysis of wireless systems allowing researchers to perform repeatable experiments. 

\fakeparagraph{Simulations.}
As WUR technology is still in its relative infancy, many simulators
have been extended for evaluating \lpwur protocols.
For benchmarking, simulators offer many advantages over testbeds. Various
network topologies such as single- and multi-hop with different traffic
patterns can be implemented and optimized with easy data collection for
extracting the metrics. Large-scale networks for scalability analysis can be
easily modeled, which otherwise would be too expensive to realize using
testbeds. Furthermore, repeatability is easily achievable in simulations. On the other hand, 
simulators are criticized for not being able to capture all
details, especially at the PHY layer, such as path loss, fading, and
interference, bringing into question the applicability of simulation results.
Nevertheless, simulation can provide valuable results.

\begin{figure}[t]
	\centering 
	\begin{subfigure}[t]{0.493\columnwidth}
		\includegraphics[width=\linewidth]{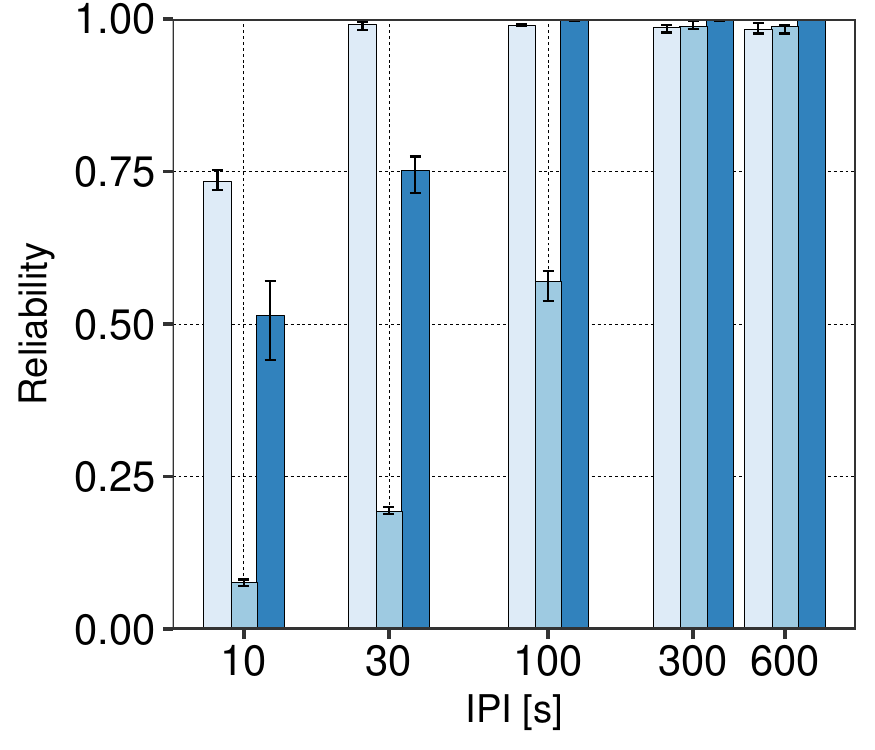}
		\caption{Network reliability.}
	\end{subfigure}
	\begin{subfigure}[t]{0.493\columnwidth}
		\includegraphics[width=\linewidth]{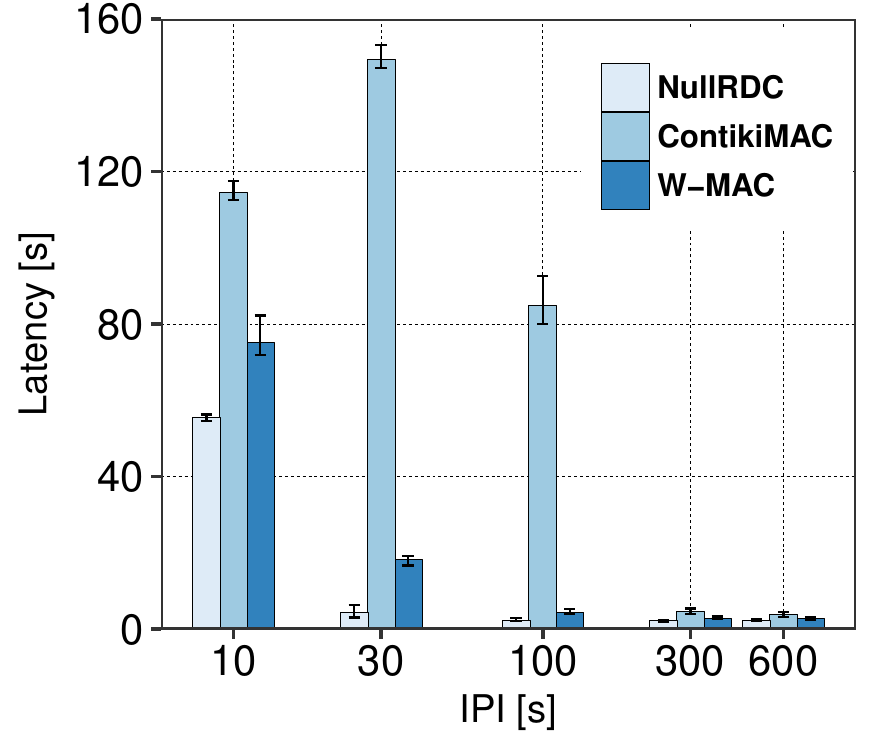}
		\caption{Data latency.}
	\end{subfigure}
	\caption{Benchmarking three different MAC protocols in simulation using WaCo.}
	\label{fig:sims}
	\vspace*{-3.0ex}
\end{figure}

\fakeparagraph{WUR simulators.}  Recent interest in WURs demands simulation
support to allow systematic exploration of this novel
technology. In~\cite{oller2016} OMNET++ extensions provide a modular
simulation model for WURs.  It employs the MiXiM framework and offers reliable
primitives for wireless signal propagation, energy consumption, and a complete
networking stack.  Similarly, GreenCastalia simulates a power model for
wake-up receivers~\cite{ spenza2015}. These simulators, however, do not offer
code portability from simulation to real system. For this, COOJA, a network
simulator widely used in the WSN community, has been augmented with
\wur~\cite{Piyare:2017}. COOJA supports node emulation for MSP430 and AVR
platforms and uses binary, deployment-ready firmware, providing the ability to
move between simulated and real experiments. It offers a full networking stack
with various signal propagation models such as Multi-path Ray-tracing and Unit
Disk Graph. Moreover, it allows simulation of multiple embedded operating
systems and is also the first open-source tool (\url {https://github.com/waco-sim}).
As an example of the ease of comparing protocols with WaCo,
Fig.~\ref{fig:sims} illustrates the benchmarking of three different MAC
protocols; wake-up radio (W-MAC), duty cycling (ContikiMAC), and always-on (NullRDC) MAC for a network of 100 nodes over Collection Tree Protocol. 
To have a fair benchmarking, the same application was run on top of all the MAC protocols with same settings while varying the network traffic.
As expected, WUR solution not only improves the network reliability but also reduces the overall latency over other MAC protocols, motivating further study of such systems using testbeds.

\fakeparagraph{Testbeds.}
There is an increasing demand for experimentally-supported results to identify
issues that cannot be captured through simulation or theory alone. This observation is reflected in the topics of the flagship conferences that increasingly encourage experimentally-driven research for validation.

Various shared and private testbeds exist, including
the  FIT IoT-LAB~\cite{fit-lab}, FlockLab~\cite{FlockLab}, and
Indriya~\cite{Indriya}. These allow scheduling experiments remotely, executing
protocols directly on hardware, as well as collecting and extracting metrics of interest from the logged data. However, none of these testbeds currently support \wur functionality. We discuss next some of the key functionalities that the testbeds need to offer and how they could be implemented for benchmarking \wurs. 

\begin{enumerate}[label=(\roman*)]
\item \emph{\wur interface}: first and foremost, testbeds must offer hardware
  with the \wur interface. One cost effective option is to support only a few
  nodes~\cite{Sutton2015}.
	
\item \emph{Experiment configurations}: the testbeds must provide experiment
  scheduling capabilities with the ability to configure a number of system
  parameters such as network topology and size, traffic load and pattern,
  experiment duration, physical layer settings for the radios including \wur
  and the main data transceiver.  As noted, these configurations
  must be clearly reported to allow comparison.

	
\item \emph{Monitoring the environment}: test facilities should provide
  information about the environmental conditions during the experiment.
  External wireless interference degrades network performance and to
  investigate this, spectrum analysis is indispensable. The testbed
  infrastructure should allow recording and replaying of the wireless traces.
  Tools such as JamLab~\cite{jamlab} are key to producing repeatable
  interference.  Temperature, instead, affects the clock oscillation of the
  devices.  TempLab~\cite{templab} offers temperature profiles for sensor
  nodes.
		
\item \emph{Data archiving and sharing}: to extend the value of
  measurements beyond a specific case study, open-source data repositories are
  necessary. This  facilitates archiving, publishing, and comparing of
  system performance data.
	
	\item \emph{Result analysis}: testbed infrastructure should be able to extract the key metrics such as power consumption, end-to-end reliability, and data latency. For instance, these metrics can be extracted  non-intrusively on testbeds using tools such as IoT-Connect~\cite{iot-connect} or D-Cube~\cite{schuss18benchmark}, avoiding the probing effects of instrumentation. 

  \end{enumerate}
Ideally, experiments should be performed in multiple testbeds to achieve statistical significance.  However, it is critical to do an ``apples to apples'' comparison in any benchmarking exercise, and failure to consider all variables can produce results that are misleading or even  erroneous. 



%% file: conclusion.tex
\section{Conclusions}
\label{sec:conclusions}

This paper offers the first steps toward \wurbench, a benchmarking framework
tailored to the unique properties of the emerging wake-up radio
technology. Solidifying this framework and encouraging it in the research
community will contribute to the solidification and wide adoption of a
technology that promises to revolutionize wireless systems.
